\documentclass[fleqn,10pt]{wlscirep}
\usepackage[utf8]{inputenc}
\usepackage[T1]{fontenc}
\usepackage{hyperref}
\usepackage{lineno}
\usepackage{pifont}
\newcommand{\cmark}{\ding{51}}
\newcommand{\xmark}{\ding{55}}
\linenumbers

\title{NuInsSeg: A Fully Annotated Dataset for Nuclei Instance Segmentation in H\&E-Stained Histological Images }

\author[1,2,*]{Amirreza Mahbod}
\author[2]{Christine Polak}
\author[2]{Katharina Feldmann}
\author[2]{Rumsha Khan}
\author[2]{Katharina Gelles}
\author[3]{Georg Dorffner}
\author[1]{Ramona Woitek}
\author[1,4]{Sepideh Hatamikia}
\author[2]{Isabella Ellinger}
\affil[1]{Research Center for Medical Image Analysis and Artificial Intelligence, Department of Medicine, Danube Private University, Krems an der Donau, 3500, Austria}
\affil[2]{Institute for Pathophysiology and Allergy Research, Medical University of Vienna, Vienna, 1090, Austria}
\affil[3]{Institute of Artifical Intelligence, Medical University of Vienna, Vienna, 1090, Austria}
\affil[4]{Austrian Center for Medical Innovation and Technology, Wiener Neustadt, 2700, Austria}

\affil[*]{corresponding author(s): Amirreza Mahbod (\texttt{amirreza.mahbod@dp-uni.ac.at)}}

\begin{abstract}
In computational pathology, automatic nuclei instance segmentation plays an essential role in whole slide image analysis. While many computerized approaches have been proposed for this task, supervised deep learning (DL) methods have shown superior segmentation performances compared to classical machine learning and image processing techniques. However, these models need fully annotated datasets for training which is challenging to acquire, especially in the medical domain.

In this work, we release one of the biggest fully manually annotated datasets of nuclei in Hematoxylin and Eosin (H\&E)-stained histological images, called NuInsSeg. This dataset contains 665 image patches with more than 30,000 manually segmented nuclei from 31 human and mouse organs. Moreover, for the first time, we provide additional ambiguous area masks for the entire dataset. These vague areas represent the parts of the images where precise and deterministic manual annotations are impossible, even for human experts. The dataset and detailed step-by-step instructions to generate related segmentation masks are publicly available at \url{https://www.kaggle.com/datasets/ipateam/nuinsseg} and \url{https://github.com/masih4/NuInsSeg}, respectively.

\end{abstract}
\begin{document}
\nolinenumbers

\flushbottom
\maketitle

\thispagestyle{empty}


\section*{Background \& Summary}

With the advent of brightfield and fluorescent digital scanners that produce and store whole slide images (WSIs) in digital form, there is a growing trend to exploit computerized methods for semi or fully-automatic WSI analysis~\cite{cui2021artificial}. In digital pathology and biomedical image analysis, nuclei segmentation plays a fundamental role in image interpretation~\cite{skinner2017nuclear}. Specific nuclei characteristics such as nuclei density or nucleus-to-cytoplasm ratio can be used for cell and tissue identification or for diagnostic purposes such as cancer grading~\cite{ skinner2017nuclear, doi:10.1177/1066896913517939, Kumar2017}. Nuclei instance segmentation masks enable the extraction of valuable statistics for each nucleus~\cite{mahbod2021cryonuseg_org}. While experts can manually segment nuclei, this is a tedious and complex procedure as thousands of instances can appear in a small patch of a WSI~\cite{Kumar2017, graham2019hover}. It is also worth mentioning that due to various artifacts such as folded tissues, out-of-focus scanning, considerable variations of nuclei staining intensities within a single image, and the complex nature of some histological samples (e.g., high density of nuclei), accurate and deterministic manual annotation is not always possible, even for human experts. The inter- and intraobserver variability reported in previous studies showing a low level of agreement in the annotation of cell nuclei by medical experts confirms this general problem~\cite{mahbod2021cryonuseg_org, monuseg}.

In recent years, many semi- and fully-automatic computer-based methods have been proposed to perform nuclei instance segmentation automatically and more efficiently. A wide range of approaches from classical image processing to advanced machine learning methods have been proposed for this task~\cite{Kumar2017, monuseg}. Up to this point, supervised deep learning (DL) methods such as Mask R-CNN and its variants~\cite{he2017mask, pmlr-v156-bancher21a}, distance-based methods~\cite{naylor2018segmentation, naylor2017nuclei} and multi encoder-decoder approaches~\cite{graham2019hover, ZHAO2020101786, 10.3389/fmed.2022.978146} have shown the best instance segmentation performances. However, to train these models, fully annotated datasets are required which is difficult to acquire in the medical domain~\cite{Kumar2017, mahbod2021cryonuseg_org, 8370747}.

A number of fully annotated nuclei instance segmentation datasets are available. These datasets were introduced for various types of staining such as Hematoxylin and Eosin (H\&E), immunohistochemical and immunofluorescence stainings~\cite{Kumar2017, kromp2020annotated, diagnostics11060967, 9446924}. The most common staining type in routine pathology is H\&E-staining. Therefore, most introduced datasets were based on this staining method. Although these datasets are valuable contributions to the research field and help researchers to develop better segmentation models, providing more annotated datasets from different organs and centers to cover more data variability is still of high importance. Table~\ref{datasets} shows the most prominent fully manually annotated H\&E-stained nuclei segmentation datasets that have been actively used by the research community in the past few years. Besides these datasets, some semi-automatically generated datasets such as PanNuke~\cite{10.1007/978-3-030-23937-4_2}, Lizard~\cite{Graham_2021_ICCV} and Hou et al. dataset~\cite{hou2020dataset} have also been introduced in the past. To generate these datasets, various approaches, such as using trained backbone models or point annotation, were exploited~\cite{graham2023conic, lin2022label, ALEMIKOOHBANANI2020101771}. However, training models based on semi-automatically generated datasets may introduce a hidden bias towards the reference model instead of learning the true human expert style for nuclei instance segmentation. 

In this work, we introduce NuInsSeg, one of the most extensive publicly available datasets for nuclei segmentation in H\&E-stained histological images. The primary statistic of this dataset is presented in the last row of Table~\ref{datasets}. Our dataset can be used alone to develop, test, and evaluate machine learning-based algorithms for nuclei instance segmentation or can be used as an independent test set to estimate the generalization capability of the already developed nuclei instance segmentation methods.

\begin{table*}[]
	\caption[]{Publicly available H\&E-stained nuclei segmentation datasets. In the table, TCGA refers to The Cancer Genome Atlas, UHCW refers to University Hospitals Coventry and Warwickshire, and MUV refers to Medical University of Vienna. The last row of the table represents the NuInsSeg dataset introduced in this work. 
	}
	\label{datasets}
	\begin{tabular}{lccccccc}
		\hline
		\textbf{dataset} & \textbf{\begin{tabular}[c]{@{}l@{}}vague \\ mask\end{tabular}} & \textbf{\# image tiles} & \textbf{\# nuclei} & \textbf{magnification} &\textbf{ \# organs} & \textbf{\begin{tabular}[c]{@{}l@{}}tile size \\ (pixels)\end{tabular}} & \textbf{source} \\
		\hline
		Kumar \textit{et al.}~\cite{Kumar2017}&\xmark & 30 & 21,623 & 40$\times$ & 7 & $1000 \times 1000$ & TCGA                \\
		MoNuSeg~\cite{monuseg}&\xmark & 44 & 28,846 & 40$\times$ & 9 & $1000 \times 1000$ & TCGA \\
		MoNuSAC~\cite{9446924}& partial & 209 & 31,411 & 40$\times$ & 4 & $81 \times 113$ to $1422 \times 2162$ & TCGA \\
		CoNSeP~\cite{graham2019hover}&\xmark & 41 & 24,319 & 40$\times$ & 1 & $1000 \times 1000$ & UHCW \\
		CPM-15~\cite{vu2019methods}&\xmark& 15 & 2,905 & 40$\times$, 20$\times$ & 2 & $400 \times 400$, $600 \times 1000$ & TCGA \\
		CPM-17~\cite{vu2019methods}&\xmark & 32 & 7,570 & 40$\times$, 20$\times$ &4& $500 \times 500$ to $600 \times 600$ & TCGA \\
		TNBC~\cite{naylor2018segmentation}&\xmark & 50 & 4,022 & 40$\times$ &1 & $512 \times 512$ & Curie Inst. \\
		CRCHisto~\cite{sirinukunwattana2016locality}&\xmark & 100 & 29,756 & 20$\times$ &1 & $500 \times 500$ & UHCW \\
		Janowczyk~\cite{janowczyk2016deep}&\xmark & 143 & 12,000 & 40$\times$ &1 & $2000 \times 2000$ & n/a \\
		Crowdsource~\cite{irshad2014crowdsourcing}& \xmark& 64 & 2,532 & 40$\times$ &1 & $400 \times 400$ & TCGA \\
		CryoNuSeg~\cite{mahbod2021cryonuseg_org}& \xmark& 30 & 7,596 & 40$\times$ &10 & $512 \times 512$ & TCGA \\
		\hline
		NuInsSeg& \cmark & 665 & 30.698 & 40$\times$ &31 & $512 \times 512$ & MUV \\
		\hline
	\end{tabular}
\end{table*}

\section*{Methods}
\subsection*{Sample preparation}
The NuInsSeg dataset contains fully annotated brightfield images for nuclei instance segmentation. The H\&E-stained sections of 23 different human tissues were provided by Associate Professor Adolf
Ellinger, PhD from the specimen collection of the Department of Cell Biology and Ultrastructural Research, Center for Anatomy and Cell Biology, Medical University of Vienna. We only obtained the stained tissue sections, not the original tissues. These images were only used for teaching purposes for a long time where no ethic votum applied.
Some of the human tissues were formaldehyde-fixed, embedded in celloidin and sectioned at $15\approx20\mu m$ (jejunum, kidney, liver, oesophagus, palatine tonsil, pancreas, placenta, salivary gland, spleen, tongue). The other human tissues were formaldehyde-fixed and paraffin-embedded (FFPE) and sectioned at $4\approx5\mu m$  (cerebellum, cerebrum, colon, epiglottis, lung, melanoma, muscle, peritoneum, stomach (cardia), stomach (pylorus), testis, umbilical cord, and urinary bladder).
Mouse tissue samples from bone (femur), fat (subscapularis), heart, kidney, liver, muscle (tibialis anterior muscle), spleen, and thymus were obtained from 8-week-old male C57BL/6J mice28. 4$\mu m$ sections of the FFPE tissue samples were stained with H\&E (ROTH, Austria) and coverslipped with Entellan (Merck, Germany).


\subsection*{Sample acquisition}
WSIs were generated with a TissueFAXS (TissueGnostics, Austria) scanning system composed of an Axio Imager Z1 (Zeiss, Oberkochen, Germany), equipped with a Plan-Neofluar 40×/0.75 objective (40× air) in combination with the TissueFAXS Image Acquisition and Management Software (Version 6.0, TissueGnostics, Austria). Images were acquired at 8-bit resolution using a colour camera (Baumer HXG40c).

\subsection*{Field of view and patch selection}
The scanning system stores individual $2048 \times 2048$ Field of Views (FOV) with their respective locations in order to be able to combine them into a WSI. Instead of using WSIs, we utilized the FOVs to generate the dataset. A senior cell biologist selected the most representative FOVs for each human and mouse WSI. From each FOV, a $512 \times 512$ pixel image was extracted by central cropping. These images were saved in lossless Portable Network Graphics (PNG) format. In total, 665 raw image patches were created to build the NuInsSeg dataset.  

\subsection*{Generation of ground truth, auxiliary, and ambiguous area segmentation masks}
\begin{figure}[ht]
	\centering
	\includegraphics[width=\linewidth]{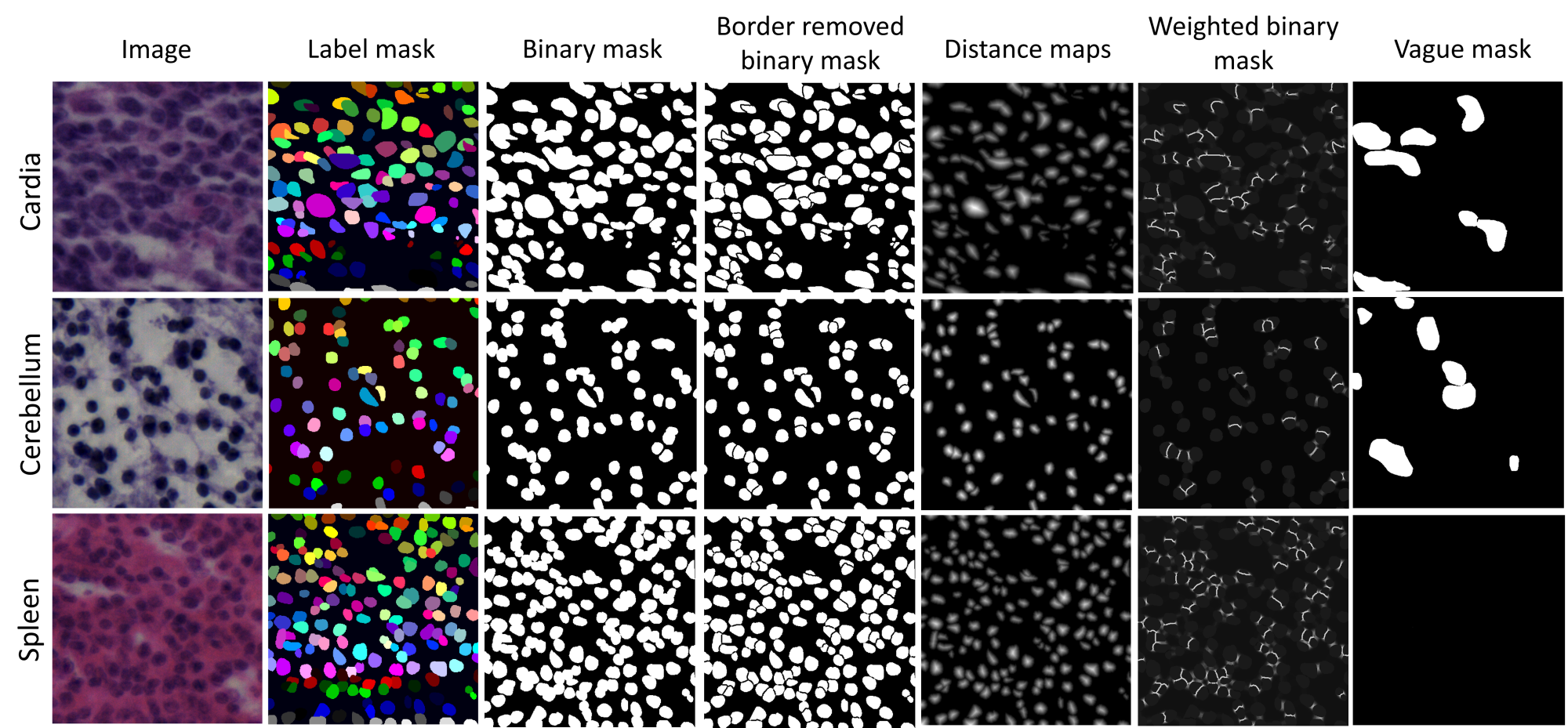}
	\caption{Example images and manual segmentation masks of three human organs from the NuInsSeg dataset. The first three columns show the original images, the labeled and the binary mask, respectively. The represented images in the fourth to sixth columns show auxiliary segmentation masks that can be beneficial for the development of segmentation algorithms. The last column shows the vague areas where accurate and deterministic manual segmentation is impossible. Some images do not contain ambiguous regions, such as the represented spleen image in the last row.}
	\label{example}
\end{figure}

We used ImageJ~\cite{schindelin2012fiji} (version 1.53, National Institutes of Health, USA) to generate the ground truth segmentation masks. We followed the same procedure suggested in ~\cite{mahbod2021cryonuseg_org} to label nuclei. We used the region of interest (ROI) manager tool (available on the Analysis tab) and the freehand option to delineate the nuclei borders. We manually draw the nuclei border for each instance until all nuclei were segmented for a given image patch. Although some semi-automatic tools such as AnnotatorJ with U-Net backbone~\cite{doi:10.1091/mbc.E20-02-0156} could be used to speed up the annotation, we stuck to fully manual segmentation to prevent any hidden bias toward the semi-autonomic annotation method. The delineated ROIs were saved as a zip file, and the Matlab software (version 2020a) was then used to create binary and labeled segmentation images (as PNG files). Besides the original raw image patches, binary and labeled segmentation masks, we also publish a number of auxiliary segmentation masks that can be useful for developing computer-based segmentation models. Auxiliary segmentation masks, including border-removed binary masks, elucidation distance maps of nuclei, weighted binary masks (where higher weights are assigned in the borders of touching objects), are published along with our dataset. The developed codes to generate these masks are available on the published GitHub repository. Moreover,   we annotated the ambiguous areas in all images of the dataset for the first time. Indicating ambiguous regions was partially provided in the test set of the MoNuSAC challenge~\cite{9745890}, but in this work, we provide it for the entire dataset. We used an identical procedure and software to create the ambiguous segmentation masks. These vague areas consist of image parts with very complex appearances where the accurate and reliable manual annotation is impossible. This is potentially helpful for in-depth analysis and evaluation of any automatic model for nuclei instance segmentation. Manual segmentation of nuclei and ambiguous areas detection were performed by three students with a background in cell biology. The annotations were then controlled by a senior cell biologist and corrected when necessary. Some example images, along with related segmentation and vague masks, are shown in Figure~\ref{example}.  

\section*{Data Records}
The NuInsSeg dataset is publicly available on a published page on the Kaggle platform (\url{https://www.kaggle.com/datasets/ipateam/nuinsseg}). The related code to generate the binary, labeled, and auxiliary segmentation masks from the ImageJ ROI files is also available on the NuInsSeg published GitHub repository \url{https://github.com/masih4/NuInsSeg}. This dataset contains 665 image patches with 30,698 segmented nuclei from 31 human and mouse organs. The organ-specific details of the generated dataset are shown in Table~\ref{details}. As shown in the table, the nuclei density in some tissues/organs (e.g., mouse spleen) is much higher in comparison to other tissues/organs (e.g., mouse muscle).

\begin{table}[]
		\caption[]{Details of the NuInsSeg dataset per human and mouse organ }
	\label{details}
	\begin{tabular}{lccccccc}
		\hline
		Organ & Type          & \#  Images & \# Nuclei & \begin{tabular}[c]{@{}c@{}}Avg. \#Nuclei\\ per image\end{tabular}   \\ \hline
		Cerebellum         & human               &12    &549     &45.8         \\ 
		Cerebrum           & human               &12    &146     &12.2        \\
	    Colon              & human               &12    &349     &29.1        \\ 
	    Epiglottis         & human               &11    &228     &20.7          \\
	    Jejunum            & human               &10    &874     &87.4          \\
	    Kidney             & human               &11    &1222    &111.1         \\
	    Liver              & human               &40    &1370    &34.3         \\
	    Lung               & human               &11    &318     &28.9          \\
	    Melanoma           & human               &12    &533     &44.4          \\
	    Muscle             & human               &9     &127     &14.1          \\
	    Oesophagus         & human               &47    &2046    &43.5          \\
	    Palatine tonsil    & human               &12    &1045    &87.1        \\ 
	    Pancreas           & human               &44    &2178    &49.5          \\     
	    Peritoneum         & human               &12    &468     &39.0          \\
	    Placenta           & human               &40    &1966    &49.2       \\ 
	    Salivary gland     & human               &44    &3129    &71.1        \\ 
	    Spleen             & human               &34    &3286    &96.7        \\ 
	    Stomach (cardia)   & human               &12    &671     &55.9         \\
	    Stomach	(pylorus)  & human               &12    &441     &36.8      \\ 
	    Testis             & human               &12    &380     &31.7        \\ 
	    Tongue             & human               &40    &1415    &35.4        \\ 
	    Umbilical cord     & human               &11    &106     &9.6        \\
		Urinary bladder    & human               &12    &400     &33.3         \\ \hline

		Bone (femur)        & mouse               &6     &757     &126.2          \\	 
		Fat (subscapularis)& mouse               &42    &549     &13.1            \\
		Heart              & mouse               &28    &738     &26.4            \\
		Kidney             & mouse               &40    &1597    &39.9            \\
		Liver              & mouse               &36    &646     &17.9           \\
		Muscle (tibialis anterior muscle)       & mouse               &28    &165     &5.9             \\ 
		Spleen             & mouse               &7     &1657    &236.7          \\ 
		Thymus             & mouse               &6     &1342    &223.7            \\ \hline
		
		All                & human               &472   &23247   &49.3            \\
		All                & mouse               &193   &7451    &38.6       &      \\ \hline
		All                & human + mouse       &665   &30698   &46.2             \\ \hline
	\end{tabular}
\end{table}

\section*{Technical Validation}
\label{technicalvalidation}

To create a baseline segmentation benchmark, we randomly split the dataset into five folds with an equal number of images per fold (i.e., 133 images per fold). We used the Scikit-learn Python package to create the folds with a fixed random state to reproduce the results (splitting code is available on the Kaggle and Github pages). Based on the created folds, we developed a number of DL-based segmentation models and evaluated their performance based on five-fold cross-validation. To facilitate to use of our dataset and developing segmentation models, we published our codes for two standard segmentation models, namely shallow U-Net and deep U-Net models~\cite{Ronneberger2015} on the Kaggle platform\footnote{\url{https://www.kaggle.com/datasets/ipateam/nuinsseg/code?datasetId=1911713}}. The model architectures of the shallow U-Net and deep U-Net are very similar to the original U-Net model but we added drop out layers between all convolutional layers in both encoder and decoder parts. Four and five convolutional blocks were used in the encoder and decoder parts of the shallow U-Net and deep U-Net, respectively. The model architecture of these two models is publicly available at our published kernels on our NuInsSeg page on the Kaggle platform. Besides these two models, we also evaluated the performance of the attention U-Net~\cite{oktay2018attention}, residual attention U-Net~\cite{oktay2018attention, He2016}, two-stage U-Net~\cite{10.1007/978-3-030-23937-4_9}, and the dual decoder U-Net~\cite{10.3389/fmed.2022.978146} models. The architectural details of these models were published in the respective articles. We performed an identical five-fold cross-validation scheme in all experiments to compare the results. For evaluation, we utilized similarity Dice score, aggregate Jaccard index (AJI), and panoptic quality (PQ) scores as suggested in former studies~\cite{Kirillov_2019_CVPR, graham2019hover, mahbod2021cryonuseg_org}. The segmentation performance of the aforementioned models is reported in Table~\ref{results}. As the results show, residual attention U-Net delivers the best overall Dice score between these models, but dual decoder U-Net provides the best average AJI and PQ scores.
Interestingly, the dual decoder model achieved the best overall PQ score in the MoNuSAC post challenge leaderborad~\cite{9745980, 9446924}, and it also achieved the best instance-based segmentation scores for the NuInsSeg dataset. It should be noted that these results can be potentially improved by using well-known strategies such as ensembling~\cite{mahbod2021pollen}, stain augmentation~\cite{li2023laplacian} or test time augmentation~\cite{wang2022fuseg} but achieving the best segmentation scores is out of the focus of this study. Instead, these results could be used as baseline segmentation scores for comparison to other segmentation models in the future, given that the same five-fold cross-validation scheme is used.
  
\begin{table}[]
		\caption[]{NuInsSeg segmentation benchmark results based on five-fold cross-validation 
	}
	\label{results}
	\begin{tabular}{lccccc}
		\hline
		Model                    & Reference                          & \# Parameters & Avg.Dice (\%) & Avg. AJI (\%) & Avg. PQ (\%) \\ \hline
		Shallow U-Net            & \cite{Ronneberger2015}             &  1.9 million  &  78.8   &  50.5        &  42.7             \\
		Deep U-Net               & \cite{Ronneberger2015}             &  7.7 million  &  79.7   &  49.4        &  40.4              \\
		Attention U-Net          & \cite{oktay2018attention}          &  2.3 million  &  80.5   &  45.7        &  36.4             \\
		Residual attention U-Net & \cite{He2016, oktay2018attention}  &  2.4 million  &  81.4   &  46.2        &  36.9             \\ 
		Two-stage U-Net          & \cite{10.1007/978-3-030-23937-4_9} &  3.9 million  &  76.6   &  52.8        &  47.2             \\ 
		Dual decoder U-Net       & \cite{10.3389/fmed.2022.978146}                     &  3.5 million  &  79.4   &  55.9        &  51.3           \\ \hline
	\end{tabular}
\end{table}

\section*{Usage Notes}

Our dataset, including raw image patches, binary and labeled segmentation masks, and other auxiliary segmentation masks, is publicly available on the published NuInsSeg page on the Kaggle platform. Step-by-step instructions to perform manual annotations and related codes to generate the main and auxiliary segmentation masks are available at our published Github repository. We also provide three kernels on the Kaggle platform to facilitate using our dataset. One kernel is devoted to explanatory data analysis (EDA), where interested researchers can visualize and explore different statistics of the NuInsSeg dataset. The other two kernels consist of related codes to perform five-fold cross-validation based on two DL-based models, namely shallow U-Net and deep U-Net, as described in the previous section.
Different Python packages were used in the coding of these kernels. To report statistics and visualize data in the EDA kernel, we mainly used Pandas (version 1.3.5) and Matplotlib (version 3.5.1) Python packages. For the DL-based model development, we mainly used Tensorflow (version 2.6.2), Keras (version 2.6.0) frameworks, and finally, for performing cross-validation, pre-and post-processing, and augmentation, Scikit-learn (version 0.23.2), Scikit-image (version 0.19.1) and Albumentation (version 1.1.0) were exploited, respectively. 

We explicitly published our dataset on the Kaggle platform, where limited free computational resources are available. Therefore, interested researchers can directly access our dataset and develop ML- or DL-based algorithms to perform nuclei instance segmentation on the NuInsSeg dataset. However, there is no limitation to downloading and saving the dataset on local systems and performing analysis using local or other cloud-based computational resources. 

It is worth mentioning that the NuInsSeg dataset can be used alone to train, validate, and test any segmentation algorithm, or it can be used as an independent test set to measure the generalization capability of already developed segmentation models.



\section*{Code availability}
The dataset and required code to generate the dataset are publicly available on Kaggle (\url{https://www.kaggle.com/datasets/ipateam/nuinsseg}) and GitHub (\url{https://github.com/masih4/NuInsSeg}), respectively.


\section*{Acknowledgements}
This work was supported by the Austrian Research Promotion Agency (FFG), No.872636. We would like to thank NVIDIA for their generous GPU donation and the TissueGnostics support team ({\url{https://www.tissuegnostics.com/}}) for their valuable advice to generate the NuInsSeg dataset. Moreover, we would like to thank Adolf Ellinger (MedUni Vienna) for providing the human tissue sections and Peter Pietschmann (MedUni Vienna) who provided the mouse samples.  

\section*{Author contributions statement}
A.M. and I.E. conceptualized the paper idea, K.G. prepared the H\&E-stained mouse sections and scanned all tissue sections, A.M., C.L., R.K., K.F., and I.E. performed annotations and controlled the segmentation masks, I.E. obtained funding, A.M conducted the experiments and reported the results, and G.D., S.H., R.W., and I.E. supervised the entire work.  All authors reviewed the manuscript. 

\section*{Competing interests} 
The authors declare no competing interests.

\bibliography{SciData}

\end{document}